\documentclass[preprint,review,12pt]{elsarticle}

\usepackage{epsfig}

\usepackage{amssymb}

\journal{Journal of Nuclear Materials}

\begin{document}

\begin{frontmatter}



\title{First principle study of intrinsic defects in hexagonal tungsten carbide}


\author{Xiang-Shan Kong$^{a}$, Yu-Wei You$^{a}$, J. H. Xia$^{a}$, C. S. Liu$^{a,\ast}\footnotetext{
*Author to whom correspondence should be
addressed. Email address: csliu@issp.ac.cn Tel: 0086-551-5591062}$
Q. F. Fang$^{a}$, G.-N. Luo$^{b}$, Qun-Ying Huang$^{b}$}


\address{$^{a}$Key Laboratory of Materials Physics, Institute of Solid
State Physics, Chinese Academy of Sciences, P. O. Box 1129, Hefei
230031, P. R. China;

$^{b}$Institute of Plasma Physics, Chinese Academy of Sciences,
Hefei 230031, P. R. China}

\begin{abstract}
The characteristics of intrinsic defects are important for the
understanding of self-diffusion processes, mechanical strength,
brittleness, and plasticity of tungsten carbide, which present in
the divertor of fusion reactors. Here, we use first-principles
calculations to investigate the stability of point defects and their
complexes in WC. Our calculation results confirm that the formation
energies of carbon defects are much lower than that of tungsten
defects. The outward relaxations around vacancy are found. Both
interstitial carbon and interstitial tungsten atom prefer to occupy
the carbon basal plane projection of octahedral interstitial site.
The results of isolated carbon defect diffusion show that the carbon
vacancy stay for a wide range of temperature because of extremely
high diffusion barriers, while carbon interstitial migration is
activated at lower temperatures for its considerable lower
activation energy. These results provide evidence for the
presumption that the 800K stage is attributed by the annealing out
of carbon vacancies by long-range migration.

\end{abstract}


\end{frontmatter}


\section{Introduction}

The International Thermonuclear Experimental Reactor (ITER) is
designed with a beryllium first wall, tungsten armor in the baffle
and divertor regions, and carbon strike point plates
\cite{Kaufmann,Pamela,Doerner}. Due to high particle and heat loads,
the wall material will be eroded and migrate as plasma impurity to
other parts. This leads to the formation of mixed material layers on
the wall surface. Previous studies have indicated that a tungsten
carbide layer could be formed during the bombardment of tungsten by
carbon impurity \cite{Eckstein,Kimura,Sugiyama}. In addition, WC is
of practical interest for engineering application due to its high
melting points, extreme hardness, good corrosion and temperature
stability \cite{Beadle,Hugosson,Isaev}.

Tungsten carbide is stable in a hexagonal structure, which consists
of alternating simple hexagonal layers of carbon and layers of
tungsten. Extensive theoretical studies have been performed to
understand the electronic structure and the origin of properties of
hexagonal tungsten carbide such as hardness, bulk module and
stability \cite{Mattheiss,Liu,Price,Suetin}. While, much less
information is available for the intrinsic defects of hexagonal
tungsten carbide, which are important for the understanding of
self-diffusion processes, mechanical strength, brittleness, and
plasticity. This may be related to the relatively low concentration
of atomic defects on the metal and the carbon sublattices. However,
the defect concentration would increase greatly in tungsten carbide
upon high-energy irradiation.

Remple \emph{et al}. \cite{Remple} firstly identified carbon and
tungsten vacancies by positron annihilation method and found the
former to be formed preferentially. Then, Ivanovskii \emph{et al}.
\cite{Ivanovskii,Medvedeva}studied the effect of tungsten and carbon
vacancies on the tungsten carbide band structure using
first-principles full-potential linear muffin-tin orbital approach.
A vacancy peak of the DOS was observed and the formation energies of
defects were calculated, i.e. 0.13 Ry (1.77 eV) and 0.20 Ry (2.72
eV) for carbon and tungsten vacancy, respectively. Thereafter,
Juslin \emph{et al}. \cite{Juslin} and Bj\"{o}rkas \emph{et al}.
\cite{Bjorkas} employed the molecular dynamic method to investigate
the intrinsic defects in hexagonal tungsten carbide. They also gave
the formation energy of tungsten/carbon vacancy and interstitial:
about 12 eV for tungsten interstitial and 3 eV for tungsten vacancy;
2.7 and 0.5 eV for carbon interstitial and vacancy, respectively.
Comparing with the results of Ivanovskii \cite{Ivanovskii}, the
difference between carbon and tungsten defect formation energies is
considerably large. Recently, Bj\"{o}rkas \emph{et al}.
\cite{Bjorkas} studied the initial state of irradiation damage in
tungsten carbides using molecular dynamics computer simulations.
They found that irradiation of tungsten carbide induce major
elemental asymmetries in defect production. This effect was
explained by the large difference between tungsten and carbon defect
formation energy. Even though certain details of the electronic
structure and formation energy of intrinsic defects in hexagonal
tungsten carbide have been addressed by aforementioned studies,
several questions on the structure and stability of native point
defects of tungsten carbide remain open, including questions about
the atomic-scale mechanisms of defect migration.

In this paper we employ first-principles calculations to study the
six possible types of native point defects of hexagonal tungsten
carbide: carbon ($\emph{V}_{C}$) and tungsten ($\emph{V}_{W}$)
vacancies, carbon (C$_{W}$) and tungsten (W$_{C}$) antisites, and
carbon ($\emph{I}_{C}$) and ($\emph{I}_{W}$) tungsten interstitials.
The formation energies of point defects are obtained, and the most
stable interstitial configurations are identified by relaxing
various possible configurations of carbon and tungsten interstitial
structure. Besides, we investigate the atomic-scale mechanisms of
isolated carbon defect migration by the nudged elastic band method.
Our calculations will provide a useful reference for further
exploration of the defect properties in tungsten carbide.

\section{Computation method}

The present calculations have been performed within density
functional theory as implemented in the Vienna \emph{ab} initio
simulation package (VASP) \cite{Kresse1,Kresse2}. The interaction
between ions and electrons is described by the projector augmented
wave potential (PAW) method \cite{Kresse3,Blohl}. Exchange and
correlation functions are taken in a form proposed by Perdew and
Wang (PW91) \cite{Perdew} within the generalized gradient
approximation (GGA). The supercell approach with periodic boundary
conditions is used to study defect properties, as well as defect
free system. The perfect supercell contains 128 atoms (64 carbon
atoms and 64 tungsten atoms) with $4\times4\times4$ unit cells. The
relaxations of atomic position and optimizations of the shape and
size of the supecell are performed with the plane-wave basis sets
with the energy cutoff of 500 eV throughout this work, which was
checked for convergence to be within 0.001 eV per atom in the
perfect supercell. Ion relaxations are performed using the standard
conjugated-gradient algorithms as implemented in the VASP code.
During the relaxations, the Brillouin zone (BZ) integrations is
achieved using a Methfessel-Pazton smearing of sigma=0.39 eV. BZ
sampling was performed using the Monkhorst-Pack scheme
\cite{Monkhorst}, with a $3\times3\times3$ \emph{k}-point mesh
centered on the Gamma point. The structural optimization is
truncated when the forces converge to less than 0.01eV.

The ground-state properties of hexagonal tungsten carbide, including
equilibrium lattice parameters and bulk cohesive energy, have been
calculated in order to compare with experimental and former
theoretical data. Results are presented in Table 1. In addition, we
obtained the bulk modulus by fitting the volume and calculated
cohesive energy to Murnaghan's equation of the state. It can be seen
from Table 1 that our results are in agreement with the experimental
and former theoretical values.

The formation energy of intrinsic defect is given by
\begin{equation}
E_{f}=E-\frac{1}{2}(n_{w}+n_{c})\mu
_{wc}-\frac{1}{2}(n_{w}-n_{c})(\mu _{w}-\mu _{c}),  \label{Eq.1}
\end{equation}
where $E$ is the cohesive energy of supercell with intrinsic defect,
$n_{w}$ and $n_{c}$ are the numbers of tungsten and carbon in
supercell, and $\mu _{wc}$, $\mu _{w}$ and $\mu _{c}$ are the
chemical potential of tungsten carbide, tungsten and carbon,
respectively. Here, $\mu _{wc}$ is taken as the cohesive energy (per
formula) of hexagonal tungsten carbide, while $\mu _{w}$ and $\mu
_{c}$ are the cohesive energy (per atom) of metallic \emph{bcc}-W
(per atom) and graphite (per atom) at their optimized geometries,
respectively.

\section{Results and discussion}

\subsection{Electronic properties}

Firstly, we investigate the electronic properties of perfect
tungsten carbide. The density of states (DOS) and valence charge
density of hexagonal tungsten carbide are shown in Fig. 1 and Fig.
2. The DOS are be calculated for $3\times3\times3$ supercells.
$13\times13\times13$ \emph{k}-grids mesh centered on the Gamma point
were obtained for Brillouin zone integrations with tetrahedron
method. For the total DOS, we can see that the Fermi level (EF) is
situated close to a minimum of the DOS that qualitatively indicates
the stability of this material. The similarity in the shapes of the
tungsten \emph{d}-projected DOS and the carbon \emph{p}-projected
DOS reflects that strong hybridization occurs between the tungsten
\emph{d} and carbon \emph{p} states. These results are consistent
with previous theoretical studies \cite{Liu,Suetin,Gaston}.

\subsection{Vacancies and antisite defects}

There are two types of vacancies in the hexagonal tungsten carbide,
namely tungsten and carbon vacancy. For the carbon vacancy, there
are six equivalent tungsten atoms as its nearest-neighbors
(W$_{Vc}^{N}$). The nearest-neighbor distance is 2.214 {\AA}.
Moreover, there are also distinct neighbor sites of carbon vacancy
aligned along the \emph{\textbf{c}} axis and in the basal plane in
carbon sublattice, denoted by C$_{Vc}^{c}$ and C$_{Vc}^{b}$,
respectively. Their distances away from the vacancy site are very
close, 2.854 and 2.932 {\AA}, respectively. Similar neighbors exist
for tungsten vacancy, and marked with C$_{Vw}^{N}$, W$_{Vc}^{c}$ and
W$_{Vc}^{b}$. Besides, there exist two types of antisite defects
that are formed by atoms located on the wrong sublattices, namely
C$_{W}$ and W$_{C}$, with a carbon atom on a tungsten site and a
tungsten atom on carbon site, respectively. These initial defect
configurations are relaxed using the DFT method described above, and
the formation energy can be obtained from the total energy of the
supercell with and without defects using Eq. (1). In addition, the
defect volume changes relative to perfect tungsten carbide are
calculated by
\begin{equation}
\Delta V/V=\frac{V_{defect}-V_{0}}{V_{0}},  \label{Eq1}
\end{equation}
where $V_{defect}$ and $V_{0}$ are the volume of supercell with and
without defect, respectively. All calculated results are summarized
in Table 2 and 3.

For the vacancy defects, we find that the nearest-neighbor atoms of
the vacancy relax outward, while other neighbors move inward. It is
this cooperative relaxation that makes the supercell volume
insensitive to the presence of vacancies at small vacancy
concentration. The volume relaxations are generally small, and the
defect volume changes relative to the perfect system are -0.2\% and
-0.4\% for $\emph{V}_{C}$ and $\emph{V}_{W}$, respectively. Similar
relaxation features have been found in other transition-metal
carbide, such as titanium carbide \cite{Tsetseris}. The formation
energy of $\emph{V}_{C}$ and $\emph{V}_{W}$ are 0.39 eV and 4.14 eV,
respectively, and the later is much larger than the former. This is
generally consistent with previous molecular dynamic
calculations\cite{Juslin,Bjorkas}, but contrasts to Ivanovskii's
results \cite{Ivanovskii}. This large difference may arise from the
difference in electronic properties between tungsten and carbon in
hexagonal tungsten carbide. In the perfect tungsten carbide,
tungsten and carbon atoms have [WC6W8] and [CW6] coordination
polyhedra, respectively. It needs six W-C bonds breaking to form a
carbon vacancy, whereas, eight W-W bonds also need to be broken to
form a tungsten vacancy. Therefore, more energy will be needed for
the tungsten vacancy formation. For the antisite defects, the
formation energy of C$_{W}$ and W$_{C}$ are significantly larger
than that of vacancy, 8.81 and 8.69 eV, respectively. This indicates
that antisites are energetically less favorable. The outward
relaxation around W$_{C}$ is considerably large, and the defect
volume change is 3.4\%. It should be pointed out that the C$_{W}$ is
not stable. The carbon atom would move away from the tungsten
vacancy site along the \emph{\textbf{c}} direction, and forms an
interstitial defect. Its formation energy is 5.29 eV.

We examine the effect of $\emph{V}_{C}$ and $\emph{V}_{W}$
imperfections on the electronic properties of tungsten carbide. Fig.
4 shows the valence charge density distribution maps for hexagonal
tungsten carbide with a carbon or tungsten vacancy. The electron
density in tungsten vacancy is much lower than that in carbon
vacancy. This explains the observation of positron annihilation
experiments, where the positron lifetime in carbon vacancies is
considerably lower than that in W vacancies. Fig. 5 and Fig. 6
present the density of states for tungsten carbide with a carbon and
tungsten vacancy, respectively. We use the same supercell and
\emph{k}-grids mesh for the DOS calculation with the defect-free
case. Comparison with the DOS of defect-free case (Fig. 1) evinces
that carbon and tungsten vacancies induce several peaks in the
valley around the Fermi level. In particular, for the DOS of
tungsten carbide with a tungsten vacancy, distinct peaks appear
about 0.3 eV below Fermi level, and 0.3 and 0.6 eV above EF. While
the most notable new feature of DOS for the case of carbon vacancy
is a peak centered about 0.2 eV above the Fermi level. Note that,
the effects of tungsten vacancy on the electronic properties are
more extensive than those of carbon vacancy. Even for this low
concentration of vacancies, the presence of the vacancy defects has
discernible effects on the electronic properties of the host
crystal.

\subsection{Self-interstitial defect}

In hexagonal tungsten carbide, there are ten different
self-interstitial sites. They are shown in Fig. 7. In naming the
self-interstitial sites, we have adopted a similar notation system
to those used in pure metal with \emph{hcp}
structure\cite{Willaime}. O is the octahedral interstitial sites
formed by equivalent three tungsten atoms and three carbon atoms.
BOC and BOW are carbon and tungsten basal plane projection of O
site, respectively. There exit two different tetrahedral
interstitial sites, i.e., TC and TW, which are formed by two
different atomic clusters: the former is composed of three carbon
atoms and one tungsten atom and the later consists of three tungsten
atoms and one carbon atom. Similarly, BTC and BTW are two different
hexahedral sites formed by two different atomic clusters: the former
is formed by three carbon atoms and two tungsten atoms, while the
later is made up of three tungsten atoms and two carbon atoms. C is
midway between the nearest-neighbor tungsten atom and carbon atom.
BCC/BCW is midway between two nearest carbon/tungsten atoms in the
dense \textbf{\emph{a}} direction in carbon/tungsten basal plane.

The configurations with carbon or tungsten atom in these
interstitial sites are relaxed using the DFT method, and the
formation energies of stable configurations are calculated using Eq.
(1). In addition, we also give the defect volume changes relative to
perfect tungsten carbide using Eq. (2). All results are summarized
in Table 4. For interstitial carbon atom, there are four stable or
metastable occupation sites, i.e., BOC, BTW, BOW, and BTC. The
formation energies of them are 3.41eV, 4.34eV, 5.00eV, and 5.67eV,
respectively. The BOC site is the energetically most favorable for
interstitial carbon atom, and others are metastable occupation
sites. This may be related to the smallest volume change, about
0.75\%, induced by the interstitial carbon atom located in BOC site.
For interstitial tungsten atom, the formation energies of all
interstitial defect configurations are considerably higher. The
preferentially occupation site is BOC for interstitial tungsten
atom, and the formation energy is 11.51eV. Moreover, there are two
complex-defects, namely $I_{C}$(BOC)+W$_{C}$ and
$I_{C}$(BTW)+W$_{C}$ (Fig. 9 (b) and (d)), consisting of an
interstitial carbon defect and a W$_{C}$ antisite defect, and which
have a little lower formation energy than interstitial tungsten
defects. We also observed two crowdion defects along the
\textbf{\emph{a}} and \textbf{\emph{c}} direction, respectively. The
formation energies of them are considerably high, about 13.65 and
15.62eV, respectively. Hence, they are energetically highly
unfavorable and can not stable in hexagonal tungsten carbide.
Comparing to the case of carbon interstitial defects, the defect
volume changes of tungsten interstitial defects are much larger.

To sum up, the formation energy of isolated tungsten defects are
much larger than that of isolated carbon defects. Anitsites and
interstitials tungsten defects are clearly energetically less
favorable. The key to understanding this large difference lies in
the covalent property of W-C bond and the large mismatch in the
covalent radii of tungsten ($r_{W}$=1.30{\AA}) and carbon
($r_{W}$=0.77{\AA}). We investigate the consequences of this
mismatch on defect formation energies for the example of the
tungsten antisite defect (Table 2). By replacing a carbon atom with
a tungsten atom the W-C bonds are replaced by W-W bonds. Without
atomic relaxation the W-W bonds are 18.4\% too small compared to
tungsten bulk. If atomic relaxation is allowed, the six neighboring
tungsten atoms move outward by about 0.22{\AA}. The atomic
relaxation is accompanied by an energy gain. However, even large
energy gain due to relaxation can not avoid that the tungsten
antisite remains energetically unfavorable. The large energy gain is
only a response to the huge internal strain which is built up by
forming this antisite. Atomic relaxation can reduce this strain;
however, it can not completely avoid it.

We also investigate the split-interstitial configuration in
hexagonal tungsten carbide. Fig. 8 shows four possible carbon
split-interstitial configurations, which are denoted by the symbols
SCC, SWC, BSCC, and BSWC, and described as follows: SCC (Fig. 8
(a)), two carbon atoms are symmetrically split in the
\emph{\textbf{c}} direction about a carbon vacant normal lattice
site; SWC (Fig. 8 (b)), a tungsten atom shares a tungsten vacant
normal lattice site with a carbon atom in the \emph{\textbf{c}}
direction; BSCC (Fig. 8 (c)), two carbon atoms are symmetrically
split in the dense \emph{\textbf{a}} direction about a carbon vacant
normal lattice site; BSWC (Fig. 8 (d)), a tungsten atom shares a
tungsten vacant normal lattice site with a carbon atom in the dense
\emph{\textbf{a}} direction. Similar configurations exit for
tungsten split interstitials, and are named by SCW, SWW, BSCW, and
BSWC, respectively. Note that, the pairs of atoms are approximately
0.5 $c_{0}$ and 2/3 $a_{0}$ apart in the split-interstitial
configuration along the \emph{\textbf{c}} and dense
\emph{\textbf{a}} direction, respectively.

Table 5 lists all calculated results of split-interstitial
configurations. For the carbon split-interstitial configurations,
starting from SCC or BSCC, the relaxations lead to form C-C dimer
configurations. Their bond lengths are 1.292 and 1.307{\AA},
respectively. The formation energies for these two dimer
configurations are 3.53 and 3.14 eV respectively, which are a little
higher/lower than that of most stable isolated carbon interstitial
defect. It should be pointed out that the dimer centre of BSCC final
configuration has shift away the carbon vacant site, as shown in
Fig. 9. For tungsten split-interstitial defects, the W-W dimer
configuration is not observed. It should be noted that the tungsten
will occupy the vacant site when a tungsten atom and a carbon atom
share a vacant site.

\subsection{Diffusion properties}

During isochronal annealing, the defects anneal out in two stages at
800 K and at 1200 K \cite{Remple}. The 800K stage is attributed by
the annealing out of carbon vacancies presumably by long-range
migration. Here, we studied the carbon vacancy and interstitial
diffusion property in tungsten carbide. In the hexagonal tungsten
carbide, there are distinct sites aligned along the
\textbf{\emph{c}} axis of the materials and in the basal plane.
Hence, we investigated two possible transition paths for carbon
defects: one path propagates parallel to the \emph{\textbf{c}} axis
and the other path is the diffusion of carbon defect in the basal
plane. The former are denoted as PC$_{V}$ and PC$_{I}$ for carbon
vacancy and interstitial, respectively; the latter are marked by
PB$_{V}$ and PB$_{I}$ for carbon vacancy and interstitial. Schematic
representations of these diffusion paths are shown in Fig. 10.

For carbon vacancy, the saddle-point structure of PC$_{V}$ is that
the carbon atom equidistant from the two carbon lattice vacant site
in the \emph{\textbf{c}} direction, i.e. BTW interstitial site (Fig.
10 (a)), while that of PB$_{V}$ is that the carbon atom seated in
the BOC as shown in Fig. 10 (c). The energy barrier of PC$_{V}$ and
PB$_{V}$ are 6.59 and 3.28 eV, respectively. This very high barrier
indicates that carbon vacancies stay more or less idle for a wide
range of temperatures, and possible redistribution or reordering of
carbon vacancies requires annealing at very high temperatures. This
result provides evidence for the presumption that the 800K stage is
attributed by the annealing out of carbon vacancies by long-range
migration. It should be pointed out that the former are much higher
than the later. This shows that vacancy prefer to diffusion in the
basal plane. For the interstitial case, the saddle-point structure
of PC$_{I}$ (Fig. 10(b)) places the interstitial carbon atom in the
BOW interstitial sites between its initial and final stable BOC
sites, and that of PB$_{I}$ is a dimer configuration as shown in
Fig. 10(d). The energy barriers are 2.20 and 0.36 eV respectively.
This lower activation energy evinces that carbon interstitial
migration is activated at lower temperatures. As same as the carbon
vacancy, interstitial carbon atom preferably diffuse in the basal
plane. In addition, the feature of high vacancy diffusion barriers
is very important for the stability of carbides, especially in the
substoichiometric (WC$_{x}$ with $x<1$) case. It implies that once
substoichiometric WC are formed, the lattice stays intact because
vacancies do not hop to other sites, unless the system are annealed
at very high temperatures and for long periods of time. In contrast,
carbon interstitial is significantly more mobile for
overstoichiometric (WC$_{x}$ with $x>1$) case. It means that the
structure of overstoichiometric WC are not stable. Our results may
give an explanation for that the overstoichiometric phase is not
observed in the phase diagram of the W-C system \cite{Kurlov}.

\section{Conclusion}

We use first-principles calculations to study the stability of point
defects and their complexes in tungsten carbide. Our calculation
results confirm that the formation energies of carbon defects are
much lower than that of tungsten defects. The outward relaxations
around vacancy are found. We identified that the BOC site is the
energetically most favorable for interstitial carbon and tungsten
atom. The C-C dimer configurations, along \emph{\textbf{c}} and
\emph{\textbf{a}} direction respectively, are also found and their
formation energy are a little higher/lower than that of most stable
isolated carbon interstitial defect. In addition, we firstly
investigated the atomic-scale mechanism of carbon defect diffusion
in tungsten carbide. Isolated carbon defect preferably diffuse in
the basal plane. The results of carbon defect diffusion show that
the carbon vacancy stay for a wide range of temperature because of
extremely high diffusion barriers, while carbon interstitial
migration is activated at lower temperatures for its considerable
lower activation energy.

\section*{Acknowledgement}





\bibliographystyle{elsarticle-num}
\bibliography{<your-bib-database>}

\begin{thebibliography}{00}

\bibitem{Kaufmann} M. Kaufmann, R. Neu, Fus. Eng. Des. 82 (2007) 521.

\bibitem{Pamela} J. Pamela, G.F. Matthews, V. Philipps, R. Kamendje, J. Nucl. Mater. 363¨C365
(2007) 1.

\bibitem{Doerner} R.P. Doerner, J. Nucl. Mater. 363¨C365 (2007) 32.

\bibitem{Eckstein} W. Eckstein, J. Roth, Nucl. Instrum. Meth. B 53 (1991) 279.

\bibitem{Kimura} H. Kimura, Y. Nishikawa, T. Nakahata, M. Oyaidzu, Y. Oyab, K.
Okunoa, Fus. Eng. Des. 81 (2006) 295.

\bibitem{Sugiyama} K. Sugiyama, K. Krieger, C.P. Lungu, J. Roth, J. Nucl.
Mater. 390¨C391 (2009) 659.

\bibitem{Beadle} K.A. Beadle, R. Gupta, A. Mathew, J.G. Chen, B. G. Willis, Thin Solid Films 516 (2008)
3847.

\bibitem{Hugosson} H.W. Hugosson, H. Engqvist, Int. J. Ref. Met. Hard Mater. 21 (2003) 55.

\bibitem{Isaev} E.I. Isaev, S.I. Simak, I.A. Abrikosov, R. Ahuja, Yu.Kh. Vekilov, M.I. Katsnelson, A.I.
Lichtenstein, B. Johansson, J. Appl. Phys. 101 (2007) 123519.

\bibitem{Mattheiss} L.F. Mattheiss, D.R. Hamann, Phys. Rev. B 30 (1984) 1731.

\bibitem{Liu} A.Y. Liu, R. M. Wentzcovitsh, M.L. Cohen, Phys. Rev. B 38 (1988) 9483.

\bibitem{Price} D.L. Price, B.R. Cooper, Phys. Rev. B 39 (1989) 4945.

\bibitem{Suetin} D.V. Suetin, I.R. Shein, A.L. Ivanovskii, J. Phys.
Chem. Solid 70 (2009) 64.

\bibitem{Remple} A.A. Rempel, R. Wurschum, and H.E. Schaefer, Phys.
Rev. B 61 (9),(2000) 5945.

\bibitem{Ivanovskii} A.L. Ivanovskii, N.I. Medvedeva, Mendeleev Commun. 1
(2001)10.

\bibitem{Medvedeva} N.I. Medvedeva, A.L. Ivanovskii, Phys. Solid State
43(2001) 469.

\bibitem{Juslin} N. Juslin, P. Erhart, P. Tr\"{a}kelin, J. Nord, K.O.E. Henriksson, K.
Nordlund, E. Salonen, and K. Albe, J. Appl. Phys. 98 (2005) 123520.

\bibitem{Bjorkas} C. Bj\"{o}kas, K. V\"{o}tler, and K. Nordlund,
Phys. Rev. B 74 (2006) 140103.

\bibitem{Kresse1}G. Kresse and J. Hafner, Phys. Rev. B 47 (1993) 558.

\bibitem{Kresse2} G. Kresse and J. Furthm\"{u}ler, Phys. Rev. B 54 (1996) 11169.

\bibitem{Kresse3} G. Kresse and D. Joubert, Phys. Rev. B 59 (1999) 1758.

\bibitem{Blohl} P.E. Bl\"{o}chl, Phys. Rev. B 50 (1994) 17953.

\bibitem{Perdew} J.P. Perdew, J.A. Chevary, S.H. Vosko, K.A.
Jackson, M.R. Pederson, D.J. Singh, and C. Fiolhais, Phys. Rev. B 46
(1992) 6671; 48 (1993) 4978(E).

\bibitem{Monkhorst} H.J. Monkhorst, J.D. Pack, Phys. Rev. B 13 (1976) 5188.

\bibitem{Pierson}H.O. Pierson, Handbook of Refractory Carbides and Nitrides: Properties
Characteristics, Processing, and Applications Noyes, Westwood, NJ,
1996.

\bibitem{Brown} H.L Brown, P.E. Armstrong, C.P. Kempter, J. Chem. Phys.
45 (1966) 547.

\bibitem{CRC}CRC Handbook of Chemistry and Physics, 85th ed., edited by D. R. Lide(CRC, Boca Raton, 2004).

\bibitem{Gaston} N. Gaston, S. Hendy, Catalysis Today 146 (2009)
223.

\bibitem{Tsetseris} L. Tsetseris, S.T. Pantelides, Acta Materialia 56 (2008)
2864.

\bibitem{Willaime} F. Willaime, J. Nucl. Mater. 323 (2003) 205.

\bibitem{Kurlov} A.S Kurlov, A.I Gusev, Russian Chemical Reviews 75(7) (2006)
617.

\end{thebibliography}






\newpage
\begin{center}
Table 1 The bulk properties and cohesive energy ($E_{c}$) of
tungsten carbide, MD and DFT mean the results obtained by molecular
dynamic simulation and density functional theory, respectively.

\begin{tabular}{ccccc}
\hline & $a_{0}$ & $c_{0}/a_{0}$ & $B_{0}($Mbar$)$ & $E_{c}($eV$)$
\\ \hline
\multicolumn{1}{l}{Experiment} & 2.907$^{a}$ & 0.976$^{a}$ & 3.29$^{b}$ & -16.68$^{c}$ \\
\multicolumn{1}{l}{MD$^{d}$} & 2.917 & 0.964 & 4.43 & -16.68 \\
\multicolumn{1}{l}{DFT$^{d}$} & 2.979 & 0.975 & 3.68 & -15.01 \\
\multicolumn{1}{l}{Present work} & 2.932 & 0.973 & 3.56 & -16.42\\
\hline\\
$^{a}$ Reference \cite{Pierson}\\
$^{b}$ Reference \cite{Brown}\\
$^{c}$ Reference \cite{CRC}\\
$^{d}$ Reference \cite{Juslin}\\
\end{tabular}
\newline

Table 2 $D$($R$)({\AA})is the distance between neighbor atom $R$
($R$=W$_{Vc}^{N}$, C$_{Vc}^{c}$, C$_{Vc}^{b}$) and defect lattice
site; $\Delta V/V$ (\%)is defect volume changes relative to the
perfect system; $E_{f}$ (eV)is the formation energy of defect in
configuration.

\begin{tabular}{cccc}
\hline
& Perfect & $V_{C}$ & W$_{C}$ \\
\hline
$D$(W$_{Vc}^{N}$) & 2.214 & 2.231 & 2.454 \\
$D$(C$_{Vc}^{c}$) & 2.854 & 2.842 & 2.817 \\
$D$(C$_{Vc}^{b}$) & 2.932 & 2.908 & 3.023 \\
$\Delta V/V$ &  & -0.2 & 3.4 \\
$E_{f}$ &  & 0.39 & 8.69\\
\hline
\end{tabular}

\mathstrut

\mathstrut

\mathstrut

Table 3 $D$($R$)({\AA}) is the distance between neighbor atom $R$
($R$=C$_{Vw}^{N}$, W$_{Vc}^{c}$, W$_{Vc}^{b}$) and defect lattice
site; $\Delta V/V$ (\%) is defect volume changes relative to the
perfect system; $E_{f}$ (eV) is the formation energy of
configuration.

\begin{tabular}{cccc}
\hline
& Perfect & $V_{W}$ & C$_{W}$ \\
\hline
$D$(C$_{Vw}^{N}$) & 2.214 & 2.311 & 2.330 \\
$D$(W$_{Vc}^{c}$) & 2.854 & 2.762 & 2.809 \\
$D$(W$_{Vc}^{b}$) & 2.932 & 2.900 & 2.911 \\
$\Delta V/V$ &  & -0.4 & -0.1 \\
$E_{f}$ &  & 4.14 & 8.81\\
\hline
\end{tabular}
\newline

Table 4 Calculation results for carbon and tungsten
self-interstitial defects in tungsten carbide. $\Delta V/V$ (\%) is
defect volume changes relative to the perfect system; $E_{f}$ (eV)
is the formation energy of defect configuration.

\begin{tabular}{ccccp{0.01in}ccc}
\hline & \multicolumn{3}{c}{Carbon interstitial defect} &  &
\multicolumn{3}{c}{ Tungsten interstitial defect} \\
\cline{2-4}\cline{6-8}
& Final~configuration & $E_{f}$ & $\Delta V/V$ &  & Final configuration & $%
E_{f}$ & $\Delta V/V$ \\ \hline
O & BOC & 3.41 & 0.75 &  & BOC & 11.58 & 4.67 \\
BOC & BOC & 3.41 & 0.75 &  & BOC & 11.58 & 4.67 \\
BOW & BOW & 5.00 & 1.82 &  & BOW & 13.48 & 4.94 \\
C & BOC & 3.41 & 0.75 &  & $I_{C}$(BOC)+W$_{C}$ & 9.73 & 3.80 \\
BCC & BOC & 3.41 & 0.75 &  & BOC & 11.58 & 4.67 \\
BCW & BTW & 4.37 & 1.96 &  & $c$-Crowdion & 13.65 & 5.32 \\
TC & BTC & 5.63 & 1.74 &  & $a$-Crowdion & 15.62 & 5.00 \\
BTC & BTC & 5.63 & 1.74 &  & $a$-Crowdion & 15.62 & 5.00 \\
TW & BTW & 4.37 & 1.96 &  & $I_{C}$(BTW)+W$_{C}$ & 10.65 & 5.21 \\
BTW & BTW & 4.37 & 1.96 &  & BTW & 15.96 & 5.94 \\ \hline
\end{tabular}

\mathstrut

\mathstrut

Table 5. Formation energy of final defect configurations obtained
from the relaxation of split interstitial configurations.  $\Delta
V/V$ (\%) is defect volume changes relative to the perfect system;
$E_{f}$ (eV) is the formation energy of defect configuration.

\begin{tabular}{cccc}
\hline & Final configuration & $E_{f}$ & $\Delta V/V$ \\ \hline
SCC & $\mathit{c}$-Dimer & 3.53 & 2.08 \\
SWC & $I_{C}$(BTC) & 5.63 & 1.74 \\
BSCC & $\mathit{a}$-Dimer & 3.14 & 1.64 \\
BSWC & $I_{C}$(BTW) & 4.37 & 1.96 \\
SCW &  $I_{C}$(BTW)+W$_{C}$ & 10.69 & 5.21 \\
SWW & $\mathit{c}$\textit{-}Crowdion & 15.62 & 5.00 \\
BSCW & $I_{C}$(BOC)+W$_{C}$ & 9.73 & 3.80 \\
BSWW & $\mathit{a}$-Crowdion & 13.76 & 5.32 \\ \hline
\end{tabular}

\end{center}

\newpage
\begin{figure}[h]
\begin{center}
\mbox{\epsfig{file=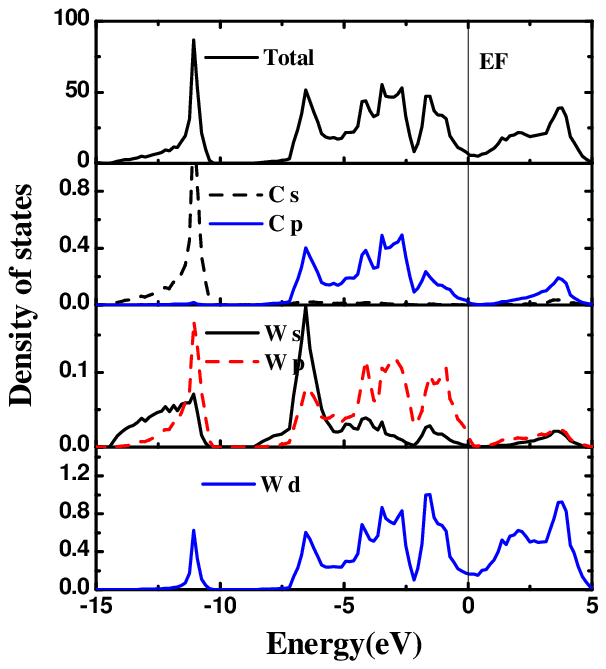}}
\end{center}
\caption{(color online) The total and partial contributions to the
DOS from orbital of different angular symmetry on carbon and
tungsten site of hexagonal tungsten carbide. Zero of energy is set
at the Fermi level. }
\end{figure}
\clearpage

\newpage
\begin{figure}[h]
\begin{center}
\mbox{\epsfig{file=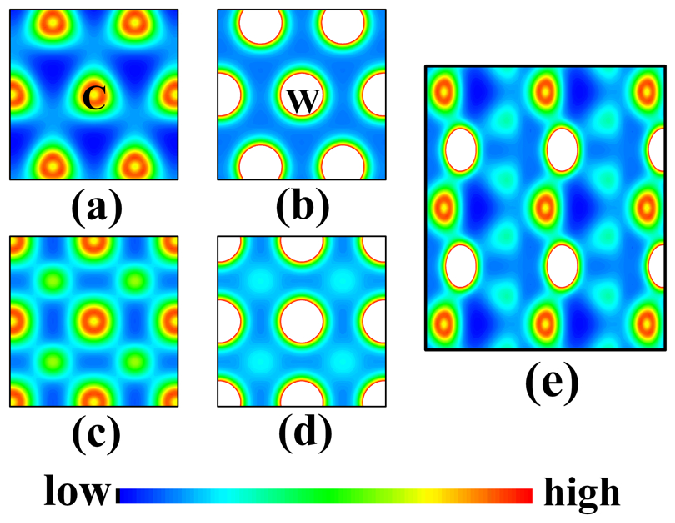}}
\end{center}
\caption{(color online) The valence charge density distribution maps
for hexagonal tungsten carbide: (a) and (b) show the valence charge
density distribution for the carbon and tungsten basal plane,
respectively; (c) and (d) present the valence charge density
distribution for carbon and tungsten (1 0 -1 0) plane, respectively;
(e) are the valence charge density in (1 1 -2 0) plane. }
\end{figure}
\clearpage

\newpage
\begin{figure}[h]
\begin{center}
\mbox{\epsfig{file=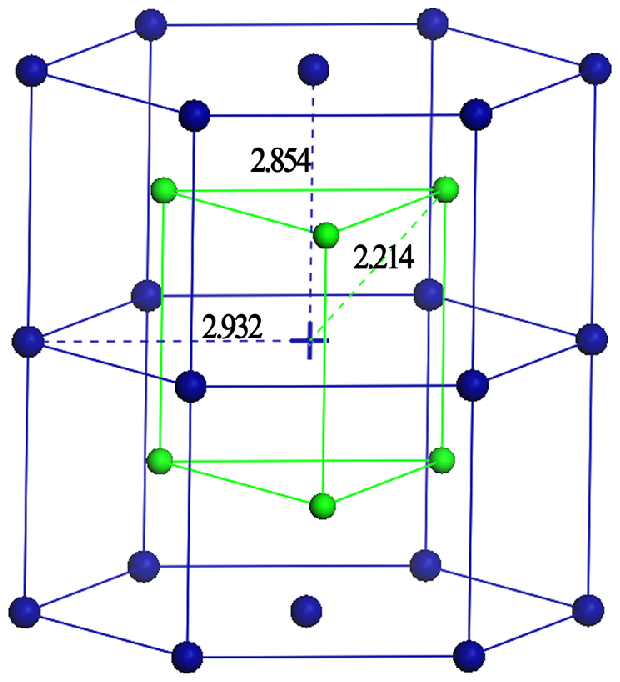}}
\end{center}
\caption{(color online)The local structure of hexagonal tungsten
carbide with a carbon vacancy. The tungsten and carbon atoms are
marked in green (white) and blue (black) balls, respectively. The
"+" denotes the carbon vacant site. The dash line is the distance
between carbon vacant and its neighbors: 2.214{\AA}, 2.854{\AA} and
2.932{\AA}. }
\end{figure}
\clearpage

\newpage
\begin{figure}[h]
\begin{center}
\mbox{\epsfig{file=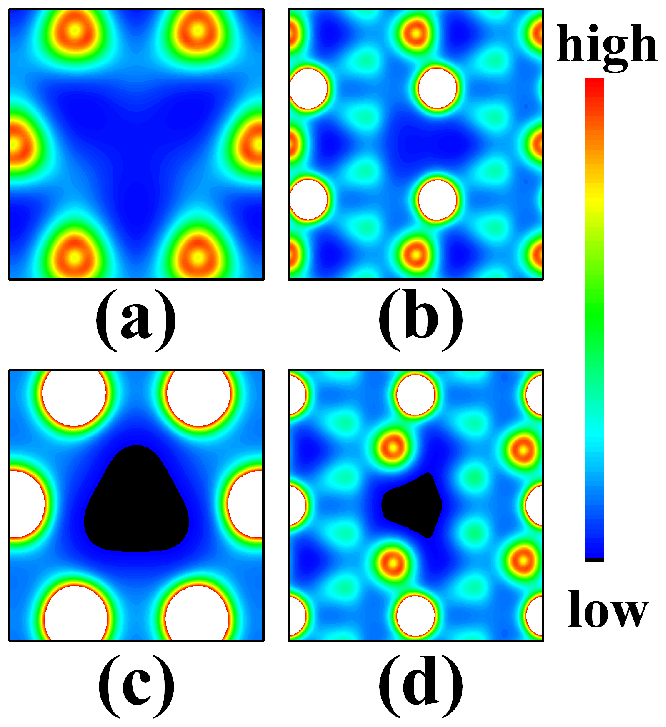}}
\end{center}
\caption{(color online) The valence charge density distribution maps
for hexagonal tungsten carbide with vacancy defect: (a) and (b) are
the valence charge density of WC with carbon vacancy in the carbon
basal plane and the (1 1 -2 0) plane, respectively; (c) and (d) are
the valence charge density of WC with tungsten vacancy in the
tungsten basal plane and the (1 1 -2 0) plane, respectively. }
\end{figure}
\clearpage

\newpage
\begin{figure}[h]
\begin{center}
\mbox{\epsfig{file=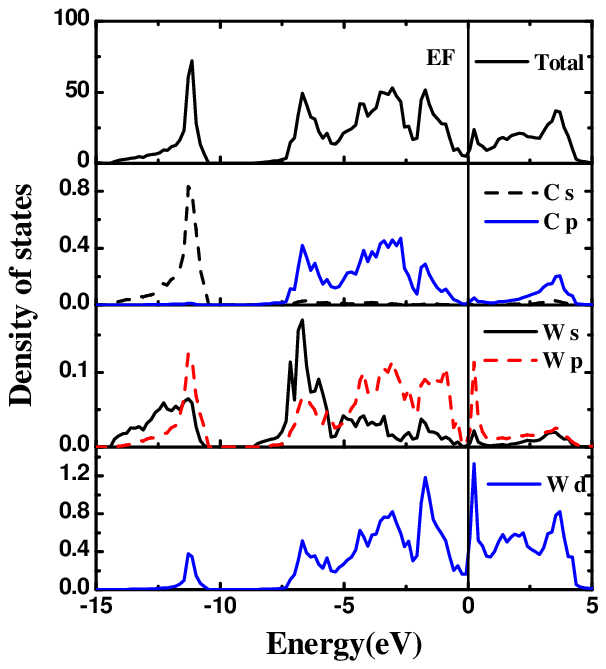}}
\end{center}
\caption{(color online)Density of states for tungsten carbide with a
carbon vacancy. }
\end{figure}
\clearpage

\newpage
\begin{figure}[h]
\begin{center}
\mbox{\epsfig{file=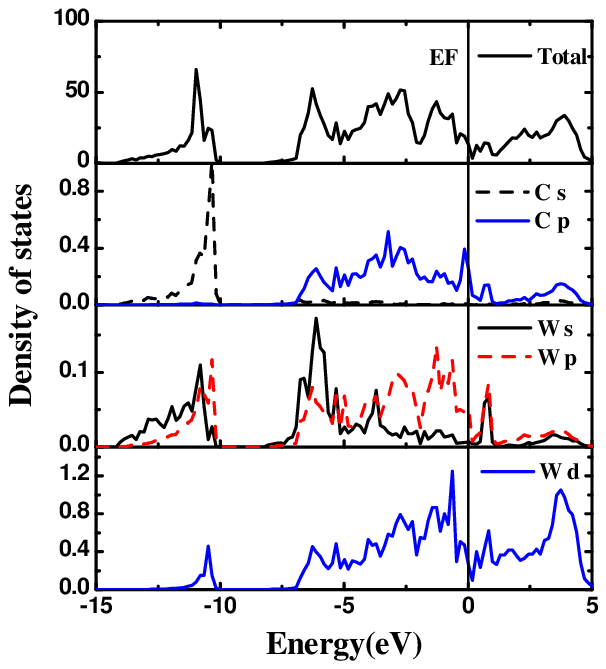}}
\end{center}
\caption{(color online)Density of states for tungsten carbide with a
tungsten vacancy. }
\end{figure}
\clearpage

\newpage
\begin{figure}[h]
\begin{center}
\mbox{\epsfig{file=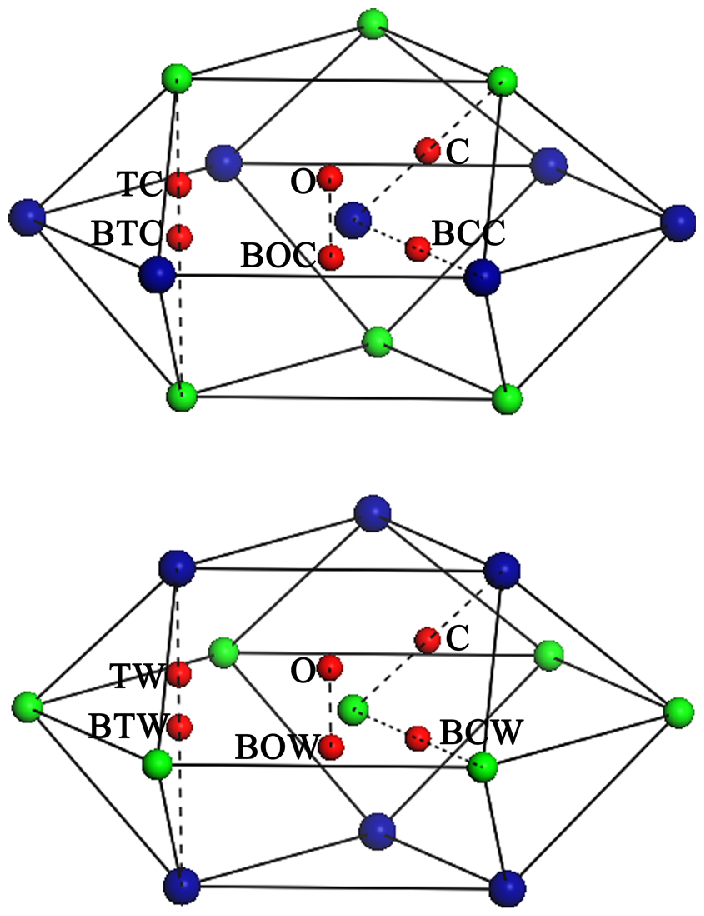}}
\end{center}
\caption{(color online) Schematic diagram of various isolated
interstitial sites in tungsten carbide. The tungsten and carbon
atoms are marked in green (white) and blue (black) balls,
respectively. The smaller red (dark gray) balls represent the
interstitial sites. }
\end{figure}
\clearpage

\newpage
\begin{figure}[h]
\begin{center}
\mbox{\epsfig{file=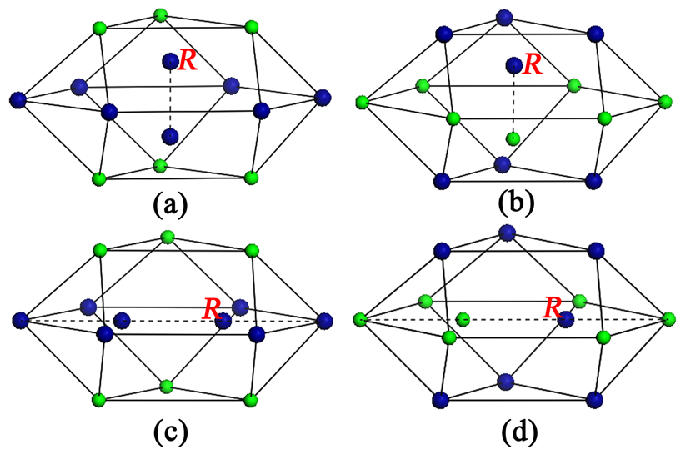}}
\end{center}
\caption{(color online) Schematic diagram of four carbon
split-interstitial configurations in tungsten carbide. The tungsten
and carbon atoms are marked in green (white) and blue (black) balls,
respectively. For tungsten split-interstitial configurations, the
carbon atom denoted by red \emph{R} would be replaced by tungsten
atom.}
\end{figure}
\clearpage

\newpage
\begin{figure}[h]
\begin{center}
\mbox{\epsfig{file=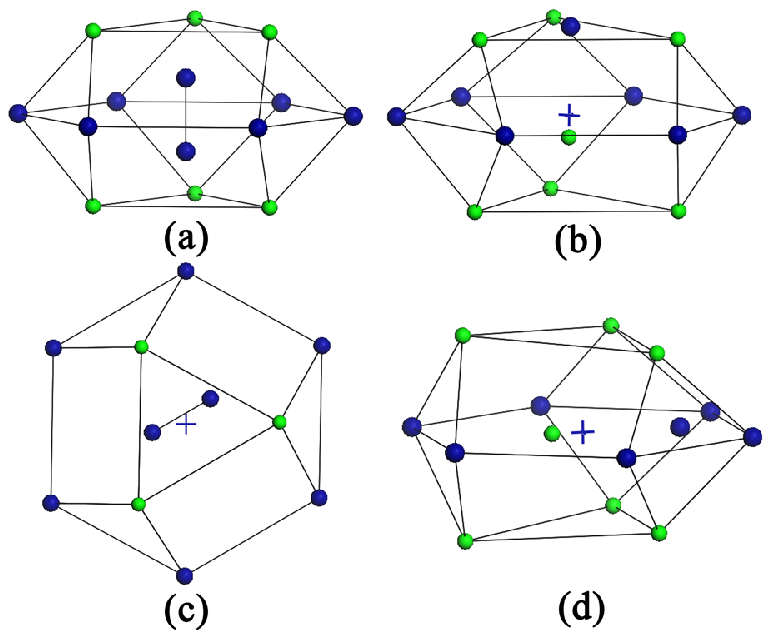}}
\end{center}
\caption{(color online) Schematic diagram of final configurations in
tungsten carbide with a interstitial carbon or tungsten atom. The
tungsten and carbon atoms are marked in green (white) and blue
(black) balls, respectively. The "+" denotes the carbon vacant site.
}
\end{figure}
\clearpage

\newpage
\begin{figure}[h]
\begin{center}
\mbox{\epsfig{file=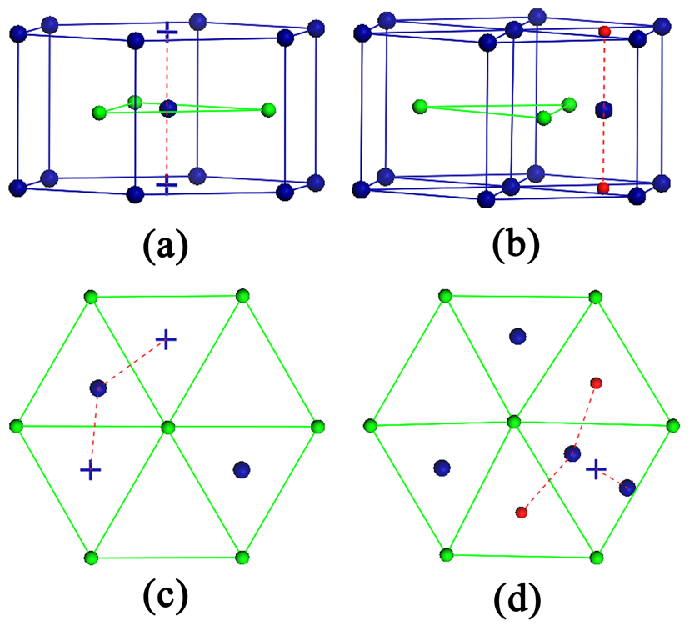}}
\end{center}
\caption{(color online) Schematic of the mechanism for carbon
vacancy and interstitial in tungsten carbide: (a) and (b) show
diffusion paths PC$_{V}$ and PC$_{I}$, respectively, while (c) and
(d) are diffusion paths PB$_{V}$ and PB$_{I}$, respectively. The "+"
denoted the initial and final vacancy site, and the small red balls
are the initial and final interstitial sites BOC. The red dash lines
are guides to the eyes for these diffusion paths. The blue and green
balls are the tungsten and carbon atoms, respectively, in the
saddle-point structure. The tungsten and carbon atoms are marked in
green (white) and blue (black) balls.}
\end{figure}
\clearpage

\end{document}